\documentclass[letterpaper, 10 pt, conference]{ieeeconf}  

\IEEEoverridecommandlockouts                              

\usepackage{xcolor}

\definecolor{pink}{rgb}{1,0.33,0.5}

\usepackage{tabularx}
\usepackage{tabulary}
\usepackage{makecell}

\usepackage{multirow}
\usepackage{subcaption}
\usepackage{booktabs}
\usepackage{threeparttable}

\let\labelindent\relax
\usepackage[inline, shortlabels]{enumitem}

\usepackage[disable]{todonotes}

\usepackage[natbib, bibencoding=utf8, citestyle=numeric, bibstyle=ieee, maxbibnames=999, maxcitenames=2, mincitenames=1, sortcites]{biblatex}
\bibliography{bibliography.bib}

\newcommand{\eg}{e.\,g. }

\newcommand{\cf}{{cf.\ }}

\usepackage[acronym, shortcuts, nohypertypes={acronym}]{glossaries}
\newacronym{AUC}{AUC}{area under the curve}
\newacronym{Bi-LSTM}{Bi-LSTM}{bidirectional long short-term memory neural network}
\newacronym{CCC}{CCC}{concordance correlation coefficient}
\newacronym{CNN}{CNN}{convolutional neural network}
\newacronym{ECG}{ECG}{electrocardiogram}
\newacronym{GBRT}{GBRT}{Gradient Boosted Regression Trees}
\newacronym{MAE}{MAE}{mean absolute error}
\newacronym{OM}{OM}{Other Runners Only Model}
\newacronym{RMSE}{RMSE}{root mean squared error}
\newacronym{RPE}{RPE}{received perception of exertion}
\newacronym{SGD}{SGD}{stochastic gradient descent}
\newacronym{sEMG}{sEMG}{surface-electromyography}

\usepackage[capitalise, nameinlink, noabbrev]{cleveref}

\title{\LARGE \bf
Fatigue Prediction in Outdoor Running Conditions using Audio Data
}

\author{Andreas Triantafyllopoulos$^{1}$, Sandra Ottl$^{1}$, Alexander Gebhard$^{1}$, Esther Rituerto-González$^{1, 2}$,
Mirko Jaumann$^{3}$,\\
Steffen Hüttner$^{3}$, Valerie Dieter$^{4}$, Patrick Schneeweiß$^{4}$, Inga Krauß$^{4}$, Maurice Gerczuk$^{1}$, Shahin Amiriparian$^{1}$,\\ and Bj\"orn W. Schuller$^{1,5,6}$
\thanks{Author correspondence to: A.\,T., {andreas.triantafyllopoulos@uni-a.de}}
\thanks{$^{1}$A.\,T., S.\,O., A.\,G., E.\,R., M.\,G., S.\,A., and B.\,S. are with EIHW – Chair of Embedded Intelligence for Health Care and Wellbeing, University of Augsburg, Germany}%
\thanks{$^{2}$E.\,R. is with the GPM – Group of Multimedia Processing, University Carlos III of Madrid, Spain}%
\thanks{$^{3}$M.\,J., S.\,H. are with HB Technologies AG, Tübingen, Deutschland}%
\thanks{$^{4}$V.\,D., P.\,S., I.\,K. are with Sportmedizin Universitätsklinikum Tübingen, Tübingen, Deutschland}%
\thanks{$^{5}$B.\,S. is also with GLAM -- Group for Language, Audio, \& Music, Imperial College, London, UK}%
\thanks{$^{6}$B.\,S. is also with audEERING GmbH, Gilching, Germany}%
\thanks{© 2022 IEEE.  Personal use of this material is permitted.  Permission from IEEE must be obtained for all other uses, in any current or future media, including reprinting/republishing this material for advertising or promotional purposes, creating new collective works, for resale or redistribution to servers or lists, or reuse of any copyrighted component of this work in other works.}%
}

\begin{document}

\maketitle
\thispagestyle{empty}
\pagestyle{empty}

\begin{abstract}
Although running is a common leisure activity and a core training regiment for several athletes, between $29\%$ and $79\%$ of runners sustain an overuse injury each year.
These injuries are linked to excessive fatigue, which alters how someone runs.
In this work, we explore the feasibility of modelling the Borg \ac{RPE} scale (range: $[6-20]$), a well-validated subjective measure of fatigue, using audio data captured in realistic outdoor environments via smartphones attached to the runners' arms.
Using \acp{CNN} on log-Mel spectrograms, we obtain a \ac{MAE} of $2.35$ in subject-dependent experiments, demonstrating that audio can be effectively used to model fatigue, while being more easily and non-invasively acquired than by signals from other sensors.
\end{abstract}
\glsresetall

\section{Introduction}
\label{sec:intro}

Running is an extremely popular past-time activity with many health benefits, and additionally plays a crucial role in several sports.
Unfortunately, running injuries are a common occurrence for professionals and novices alike~\citep{saragiotto2014main}.
Understanding the cause of those injuries and detecting behaviours that lead to them has been a major source of study in sports science over the years~\citep{nielsen2012training}.
Moreover, a holistic characterisation of runner behaviour can be highly beneficial for longtime athletes who wish to improve their performance through training, as well as amateurs who wish to avoid injuries while leveraging the benefits of a healthy lifestyle.

One particularly poignant danger is that of overuse injuries, which impacts $29\%$ to $79\%$ of runners on a yearly basis~\citep{van2007incidence}.
Overuse injuries are caused by repetitive movement during running which stresses bodily structures (such as tendons or bones), sometimes beyond their tolerance point.
These injuries have furthermore been linked to asymmetries between the left and the right side while running, which are in turn linked to a runner's \emph{fatigue} levels~\citep{schutte2016fatigue, op2018fatigue}.
By early and accurately predicting an individual's fatigue levels, we can inform their running behaviour in an attempt to prevent overuse injuries.
Furthermore, it can be used to improve performance through personalised training regiments.

Previous studies on automatically predicting runner fatigue have primarily focused on biomechanical and heart-rate sensors~\citep{op2018fatigue, gholami2020fatigue, guan2021sports}.
Typically, these studies rely on accelerometer data from wearable sensors attached to various parts of a runner's body (\eg, their arms or legs), which is often combined with other physiological data.
While the feasibility of using such sensors to predict user fatigue has long been established, using such sensors in a real-world setting could be highly impractical, as users would need to buy such sensors and learn to operate them properly.
This constitutes a high entry barrier for amateur users who wish to benefit from such technology but do not have the budget, or the patience, for such sophisticated equipment.

Moreover, several prior works have focused on `laboratory' conditions, with data collected in indoor environments using treadmills~\citep{gholami2020fatigue, guan2021sports}, whereas others that went beyond that monitored people in running tracks~\citep{op2018fatigue}.
However, another large amount of people opt to run outdoors, where some conditions are uncontrollable and over several different surfaces (\eg, asphalt, gravel, or concrete).
This limits the real-world applicability of algorithms developed for controlled conditions.

We attempt to mitigate those shortcomings by collecting a new dataset of $48$ runners in outdoor conditions.
This enables to monitor their behaviour under realistic circumstances, thus the generalisability of our results.
Furthermore, in addition to the usual wearable sensors used in other studies, we collect audio data through smartphones attached to the runners' bodies using off-the-shelf armbands (\cf \cref{sec:data}).
In this work, we are primarily interested in the feasibility of using this audio data collected to predict self-reported fatigue in realistic, outdoor conditions -- a feat which would widely expand the target reach of running monitoring applications.

The remainder of this paper is organised as follows.
\cref{sec:related_work} outlines the state of the art regarding fatigue prediction.
In \cref{sec:data}, we introduce our dataset and data collection process.
\Cref{sec:setup} describes the audio-based method used to predict results.
\Cref{sec:results} and \cref{sec:conclusion} contain the results and conclusion of our work, respectively.

\section{Related Work}
\label{sec:related_work}

Several prior works examine running-associated fatigue.
In general, the usage of wearable sensors attached to the runner's body is a common way of recording and monitoring corresponding features. 
Some of the most commonly used sensors are accelerometers, biomechanical, and physiological sensors~\citep{op2018fatigue,guan2021sports,khan2019novel,leduc2020convergent}. 

For example, \citet{strohrmann2012monitoring} monitored kinematic changes resulting from fatigue in running and argued that recordings from treadmills do not always generalise well to outdoor running.
In a similar setting, \citet{eskofier2012embedded} recorded heart rate, heart rate variability, running speed, stride frequency, and biomechanical data during a free one-hour outdoor run. 
These features were used to train classifiers which could distinguish between two levels of fatigue with an accuracy of $88.3\,\%$.

\citet{op2018fatigue} analysed whether ML can be properly utilised to predict the \ac{RPE} using data from inertial sensors placed on participants running outdoors.
They predicted the \ac{RPE} value over a longitudinal dataset of $29$ runners~\citep{op2018fatigue}, and achieve a best \ac{MAE} of $2.03$ using arm sensors. 

\citet{guan2021sports} examined inertial and physiological sensors to collect \ac{ECG}, acceleration, and angular velocity signals. 
An attention \ac{Bi-LSTM} was employed to classify three levels of sports fatigue achieving a recognition accuracy of $80.55\%$.

\citet{khan2019novel} proposed a novel method for classifying running fatigue by applying change-point segmentation. 
The experiments were performed during an incremental treadmill running-test at which \ac{sEMG} was utilised to non-invasively monitor the lactate concentration in blood. 
They trained three separate random forest classifiers to predict three different levels of fatigue and achieved an average \ac{AUC} score of over $0.8$.

Recently, textile sensors have also been considered.
\citet{gholami2020fatigue} exploited this type of sensor to continuously monitor the kinematics during running. The sensor was applied to five female participants during running, and a stacked random forest model was trained to predict the perceived exertion levels. 
They report a \ac{RMSE} value of $0.06$ and a coefficient of determination ($R^2$) of $0.96$ in participant-specific scenarios.


\section{Dataset}
\label{sec:data}

This work is based on the KIRun dataset, a newly-introduced running dataset consisting of $48$ runners in ‘in-the-wild’ outdoor conditions, comprising a total of $185$ running sessions ($[1-5]$ per subject) of each approximately $45$ minutes long.
Runs took place in different locations across Germany.
The distribution of age ranges and sex of the runners is detailed in \cref{tab:runners_stats}. 
Each running session consists of different modalities of gathered data: audio signal, heart rate, biomechanical knee sensor and foot sensor data.
Ethical approval was obtained from the ethics committee of the University Hospital Tuebingen. 
The study was registered in the German clinical trial register (DRKS00025380).
In this work, we focus exclusively on the audio data.

\begin{table}[t]
\centering
\caption{Distribution of runners per age range and sex.}
\label{tab:runners_stats}
\begin{tabular}{@{}c|cc|c@{}}
\makecell{\textbf{Age range} \\ (Years)}                 & \makecell{\textbf{(M)ale} \\ (\#)} & \makecell{\textbf{(F)emale} \\ (\#)}                  & \makecell{\textbf{Total} \\ (\#)} \\ \midrule
21 - 30                             & 5             & \multicolumn{1}{c|}{7}           & 12             \\
31 - 40                             & 8             & \multicolumn{1}{c|}{8}           & 16             \\
41 - 50                             & 2             & \multicolumn{1}{c|}{4}           & 6              \\
51 - 60                             & 6             & \multicolumn{1}{c|}{8}           & 14             \\ \midrule
\multicolumn{1}{l|}{\textbf{Total} (\#)} & 21   & \multicolumn{1}{c|}{27} & \textbf{48}   
\end{tabular}
\end{table}


During the running session, the audio recordings are captured by smartphones attached to the runners' bodies using armbands. 
Five different smartphone models were used for capturing such data throughout all sessions, all with an operating system higher than iOS 14. 
The audio signals recorded were captured at a sampling frequency of $16$\,kHz and a bitrate of $16$ bits in stereo sound, and were converted to a single channel to facilitate their analysis.

Every $3 - 5$ minutes, a smartphone app guiding the running session asked questions aloud through the phone speaker about the runner's general state, and their responses were subsequently recorded. 
Specifically, the subjects were asked to rate their \emph{fatigue} in a range from $6$ to $20$, their  \emph{wellbeing} ranged from $-5$ to $5$, and the \emph{surface} on which they were running, given as a free answer.

\begin{figure}[t]
    \centering
    \begin{subfigure}[b]{.49\linewidth}
    \includegraphics[width=\textwidth]{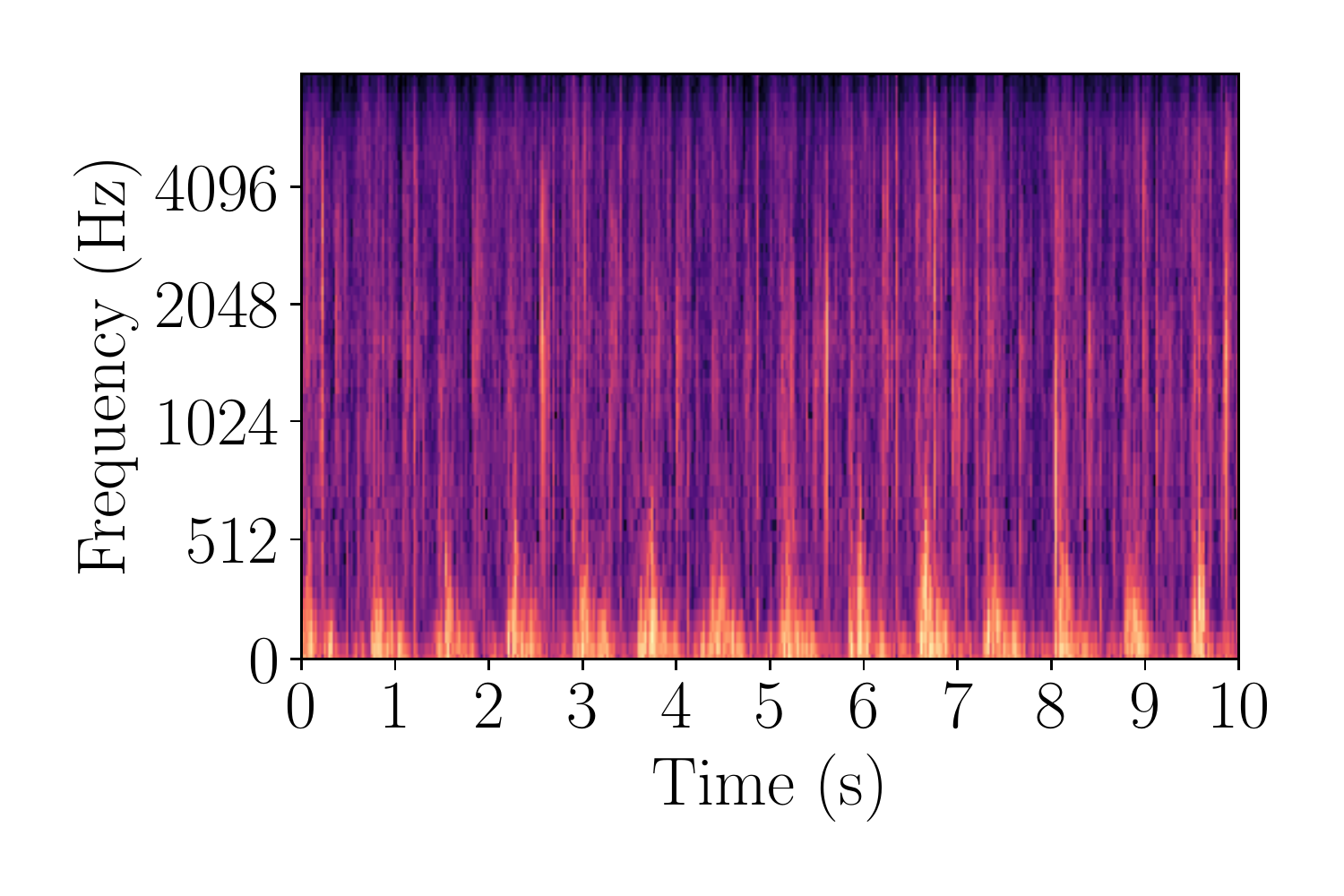}
    \end{subfigure}
    \begin{subfigure}[b]{.49\linewidth}
    \includegraphics[width=\textwidth]{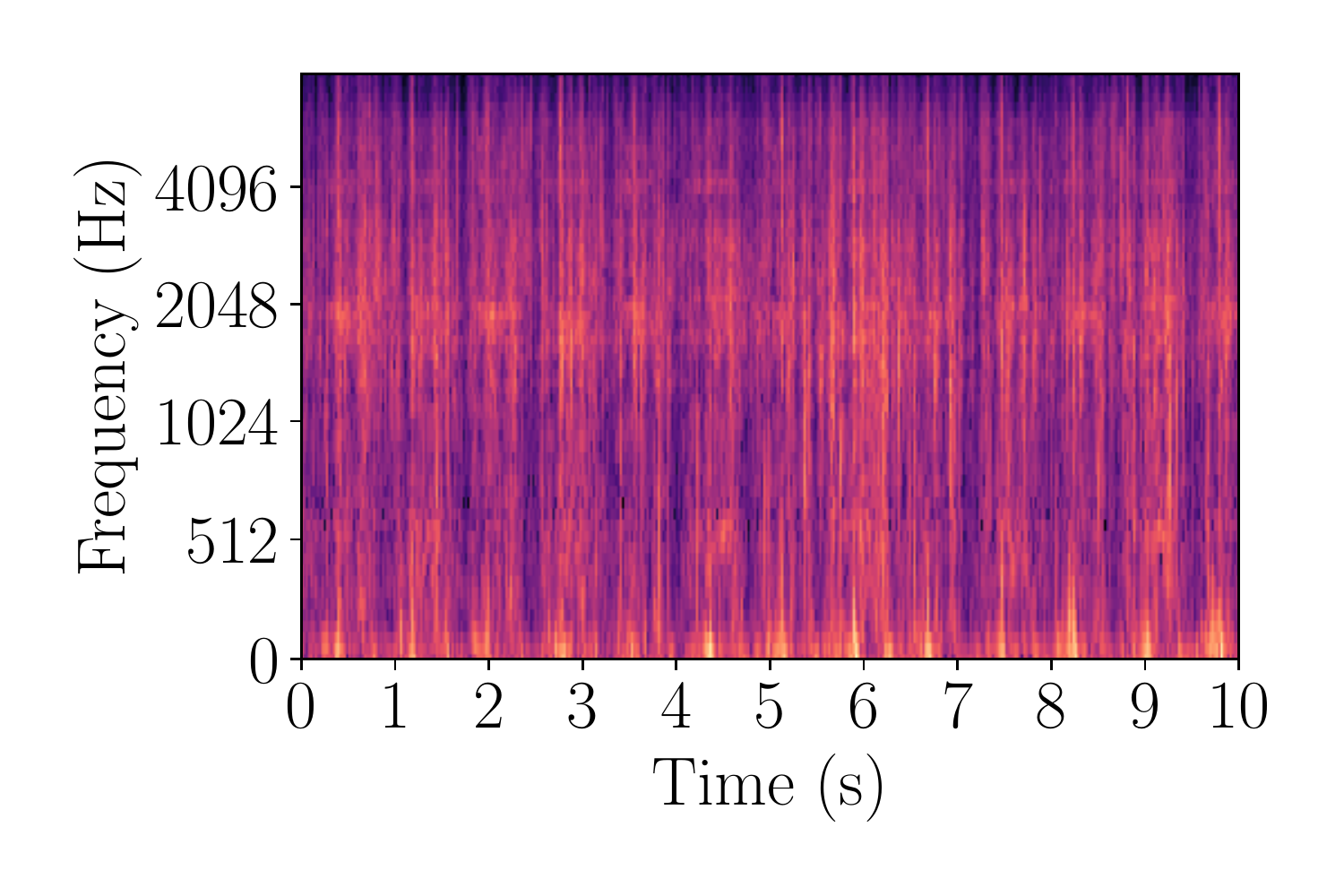}
    \end{subfigure}
    
    \caption{
    Two example log-Mel spectrograms ($3$ seconds) from the same runner with $\text{RPE} = 10$ (left) and $\text{RPE} = 17$ (right).
    Runner is $49$ years old, female, with $~20$ years of running experience and a BMI of $18.4$.
    }
    \label{fig:spectrograms}
    \vspace{-1em}
\end{figure}

The runners' answers correspond exactly to their physical condition for only short time-window around the moment they gave them.
In this work, we make the assumption that those values remained constant for the $30$ seconds surrounding the answer time.
We we segment the continuous recordings to contain data $15$ seconds prior and $15$ seconds after the answers.
Each audio segment was then labelled according to the answers of the runner, and therefore has one unique \emph{wellbeing}, \emph{fatigue}, and running \emph{surface} label.
This resulted in $15$ hours of processed audio data, out of the $133$ hours of the original raw audio data.

\Cref{fig:spectrograms} show an example of two log-Mel spectrograms from the same runner, each of them covering ten seconds, during different \ac{RPE} levels.
The spectrogram with the lower \ac{RPE} (on the left) shows regularly occurring energy bursts in lower frequencies(~$[0-100]$\,Hz), corresponding to their steps.
However, the spectrogram on the right, corresponding to a higher \ac{RPE}, additionally presents energy in medium frequencies (~$2000$\,Hz) corresponding to heavy breathing. 

\begin{figure}[t]
    \centering
    \includegraphics[width=.89\linewidth]{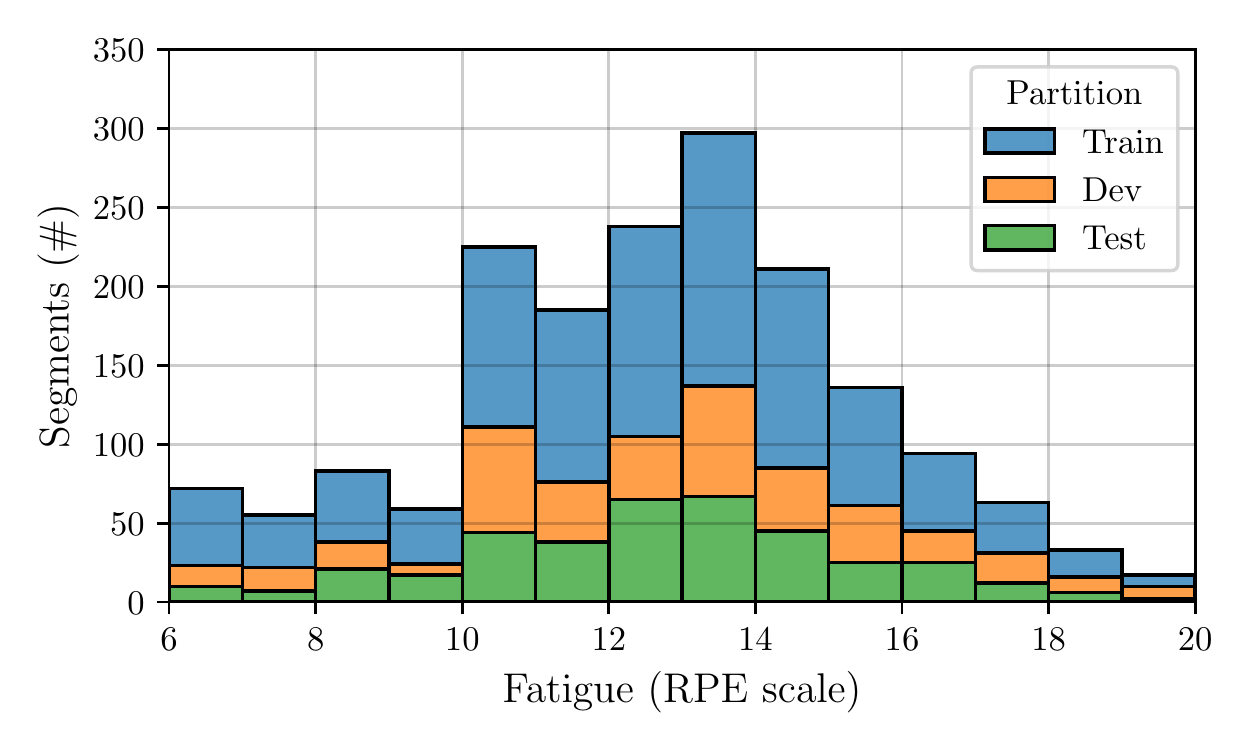}~%
    
    \caption{
    Runner fatigue distribution in the KIRun dataset.
    }
    \label{fig:tasks}
\end{figure}

Finally, we split the data in three partitions (training, development, and test) such that no running sessions from training and development were present in the test partition. 
The partitions are subject dependent, containing $56$\,\%, $23$\,\% and $21$\,\% of the data, respectively, and their distribution is shown in \cref{fig:tasks}.
Most segments fall into the $[10-16]$ scale, thus showing signs of moderate fatigue.

\section{Experimental Setup}
\label{sec:setup}
As discussed in \cref{sec:intro}, we are interested in using the audio signal alone to predict each runner's \emph{fatigue}.
This audio signal was captured by smartphone microphones as described in \cref{sec:data}.

We use the CNN14 architecture introduced by \citet{kong2019panns}, which is based on the VGG architecture design.
CNN14 consists of $6\times2$ convolutional blocks, each followed by max pooling and a dropout with probability $0.2$.
Convolution layers use a $3\times3$ kernel and a stride of $1\times1$, whereas all max pooling layers use a stride of $2\times2$, with the exception of the last one which uses $1\times1$.
Following the last block, features are pooled over the time dimension using both mean and max pooling and fed into two linear classification layers.

We investigated two experiment variants.
We first trained CNN14 from a random initialisation, which gives us the baseline performance that this architecture can achieve, and subsequently initialised the model with publicly-available pre-trained weights\footnote{https://github.com/qiuqiangkong/audioset\_tagging\_cnn}.
These were obtained by the original authors of CNN14 after training on AudioSet~\citep{gemmeke2017audio} for audio tagging, where we substitute the last (linear) classification layer with a randomly initialised one for our task.
This allowed us to utilise the power of learnt representations -- a major cornerstone of recent deep learning successes.
The two variants are denoted as \textbf{CNN14-random} and \textbf{CNN14-pretrained}, respectively.

As features, we used log-Mel spectrograms extracted over the entire $30$ second segments.
The spectrograms were computed with $64$ Mel bins, a window size of $32$\,ms, and a hop size of $10$\,ms.
Our sampling rate for all experiments was $16$\,kHz.
All models were trained for $50$ epochs with a batch size of $24$ using an \ac{SGD} optimiser with a learning rate of $0.001$, a Nesterov momentum of $0.9$, and a weight decay of $0.0001$.
As loss function we used \ac{CCC}, which is commonly utilised for auditory intelligence (regression) tasks with self-reported labels~\citep{trigeorgis2016adieu, triantafyllopoulos2021multistage}.
The best-performing epoch was selected based on development set performance.
\section{Results}
\label{sec:results}

The results of our experiments are presented in \cref{table:results}.
As performance metrics, we use \ac{CCC} and mean average error \ac{MAE}.
A maximum \ac{CCC} of $.287$ and minimum \ac{MAE} of $2.35$ is obtained for \emph{fatigue} on the Borg \ac{RPE} scale ranging $[6-20]$ using CNN14-pretrained, while results obtained with random initialisation are substantially worse, in line with previous works showing the effectiveness of transfer learning~\citep{gerczuk2021emonet}.

\begin{table}[]
    \centering
    \caption{
    Fatigue prediction \acl{MAE} and \acl{CCC} (\ac{MAE}/\ac{CCC}) test set results on the KIRun dataset.
    }
    \label{table:results}
    \begin{tabular}{c|cc}
        \toprule
        \thead{Model} & \thead{\ac{MAE}} & \thead{\ac{CCC}} \\
        \midrule
        \thead{CNN14-random} & 3.48 & .208\\
        \thead{CNN14-pretrained} & \textbf{2.35} & \textbf{.287}\\
        \bottomrule
        \end{tabular}
\end{table}

\begin{figure}[t]
    \centering
    \includegraphics[width=.89\linewidth]{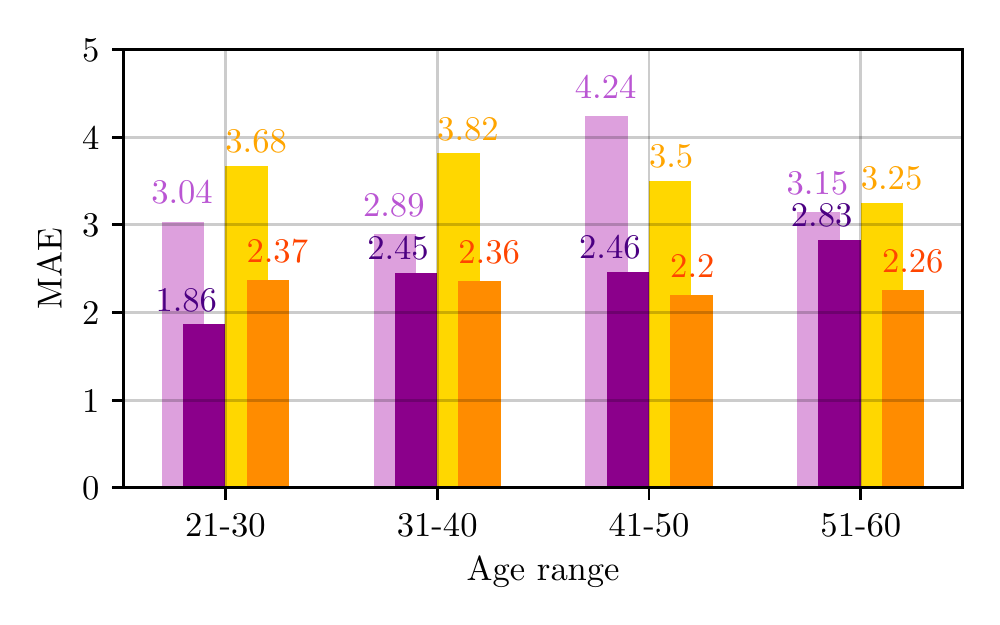}~%
    
    \caption{
    Age- and sex-based stratification of MAE results. 
    Violet refers to male, and orange refers to female. Light colours refer to CNN14-random and dark to CNN14-pretrained.
    }
    \label{fig:diversity}
\end{figure}

Furthermore, we investigated the performance of our models from the perspective of fairness~\citep{mitchell2019model}.
\Cref{fig:diversity} shows \ac{MAE} results per age range and sex.
We observe that the age ranges containing $[21-50]$ show a more stable performance whereas the range $[51-60]$ shows the highest \acp{MAE}.
This is despite the fact that this age range is well represented in the data (\cf \cref{tab:runners_stats}).
Surprisingly, even though the age range between $[41-50]$ is the most underrepresented one (\cf \cref{tab:runners_stats}), it still yields very low \ac{MAE}.
Moreover, with respect to biological sex, our model perform almost the same for male and female runners.
This is despite the fact that our data is largely biased towards females ($27$ female runners vs $21$ male).
This indicates that relative data quantity may not be the only factor causing performance imbalances.
Finally, we note that while CNN14-pretrained performance is overall better for females in most age groups, the reverse is true for the age group $[21-30]$, where females show an \ac{MAE} of $2.37$ compared to $1.86$ for males.

Another interesting pattern is the difference of behaviour between CNN14-random and CNN14-pretrained.
For some particular age groups, the two models show very different behaviour.
For example, for the age group $[31-40]$ CNN14-random shows a much larger \ac{MAE} for females, but the performance of CNN14-pretrained is almost the same for both sex groups.
This shows that pre-trained models not only improve absolute performance, but might also change model behaviour across different strata of the dataset -- which is an unwanted side-effect of the underspecification phenomenon observed in ML architectures~\citep{triantafyllopoulos2021fairness}.

\begin{figure}
    \centering
    \includegraphics[width=.99\linewidth]{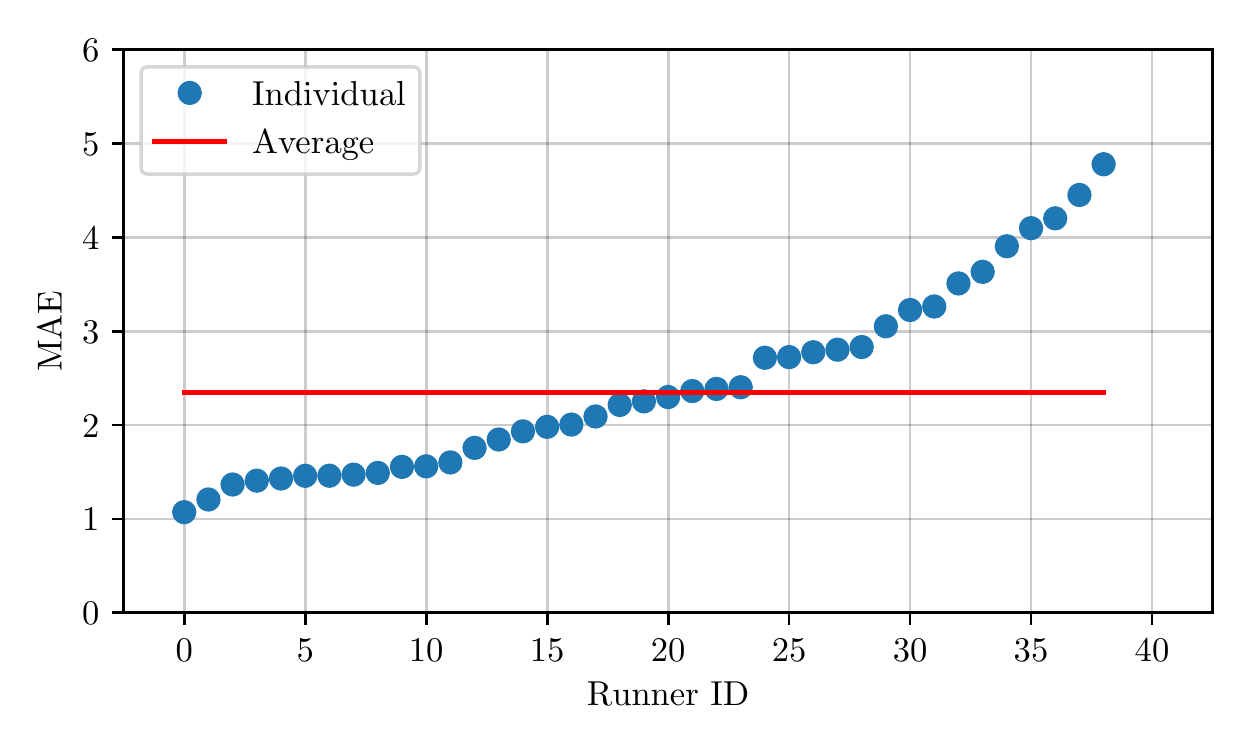}
    \caption{
    Individual-based vs average \ac{MAE} performance.
    \ac{MAE} scores computed using the samples of each runner separately (blue).
    Global \ac{MAE} computed over all samples (red).
    Runners sorted from best to worst performance.
    }
    \label{fig:individual}
    \vspace{-1em}
\end{figure}

We end with a discussion on individual-level performance.
\Cref{fig:individual} shows individual fatigue \ac{MAE} results obtained with CNN14-random and CNN14-pretrained for each runner in the test set.
The figure illustrates that while a global \ac{MAE} of $2.35$ is obtained, the performance of several runners is worse, reaching \ac{MAE} scores up to $5$.
This is indicative of individual differences in running behaviour that can be tackled with personalisation approaches that adapt to each runner~\citep{triantafyllopoulos2021deep}.
Such approaches would balance the individual runner characteristics and potentially mitigate the fairness gap as well.

\section{Conclusion}
\label{sec:conclusion}
In this work, we investigate audio-based fatigue prediction from runners in outdoor environments using \acp{CNN}.
We achieve an \ac{MAE} of $2.35$ audio data alone for fatigue prediction on the Borg \ac{RPE} scale (ranging $[6-20]$), which fares favourably against slightly more accurate approaches that use substantially harder-to-acquire sensors~\citep{op2018fatigue}.
Moreover, our intersectional fairness analysis reveals that performance differs between age groups and sex combinations, and that individual-level performances are important.
Future work could be targeted at multimodal approaches that integrate all available sensors~\citep{triantafyllopoulos2021multistage}, as well as personalisation techniques that are tailored to individual running patterns~\citep{triantafyllopoulos2021deep}.

\addtolength{\textheight}{-12cm}   




\section*{ACKNOWLEDGMENT}

This work was funded from the DFG's Reinhart Koselleck project No.\ 442218748 (AUDI0NOMOUS), the Zentrales Innovationsprogramm Mittelstand  (ZIM) grant agreement No.\ 16KN069455 (KIRun), an EB-FPU Grant 
from the Spanish Ministry of Universities (MIU), and a Short-Term Grant from the DAAD (Deutscher Akademischer Austauschdienst).

\section{\refname}
\printbibliography[heading=none]

\end{document}